\documentclass[conference]{IEEEtran}
\IEEEoverridecommandlockouts
\usepackage{amsmath,amssymb,amsfonts}
\usepackage{algorithmic}
\usepackage{soul}
\usepackage{subcaption}
\usepackage{graphicx}
\usepackage{textcomp}
\usepackage{xcolor}
\usepackage{acronym}
\usepackage[
backend=bibtex,
style=numeric-comp, sorting=none, isbn=false,
eprint=false, doi=false, url=false, arxiv=abs]{biblatex}

\definecolor{blueflat}{RGB}{41,128,185}  
\usepackage{hyperref}
\hypersetup{colorlinks,linkcolor={black},citecolor={blue},urlcolor={blue}}  
\begin{document}
\title{Heterogeneous Population Encoding for Multi-joint Regression using sEMG signals}
\author{\IEEEauthorblockN{Farah Baracat}
\IEEEauthorblockA{\textit{Institute of Neuroinformatics} \\
\textit{University of Z{\"u}rich and ETH Z{\"u}rich}\\
Z{\"u}rich, Switzerland \\
fbarac@ini.uzh.ch}
\and
\IEEEauthorblockN{Luca Manneschi}
\IEEEauthorblockA{\textit{School of Computer Science} \\ 
\textit{University of Sheffield }\\ 
Sheffield, S10 2TN, UK\\
l.manneschi@sheffield.ac.uk}
\and
\IEEEauthorblockN{Elisa Donati}
\IEEEauthorblockA{\textit{Institute of Neuroinformatics} \\
\textit{University of Z{\"u}rich and ETH Z{\"u}rich}\\
Z{\"u}rich, Switzerland \\
elisa@ini.uzh.ch}
}
\acrodef{sEMG}[sEMG]{Surface Electromyography}
\acrodef{RMSE}[RMSE]{root mean square error}
\acrodef{MAE}[MAE]{mean absolute value}
\acrodef{HMI}[HMI]{human-machine interface}
\acrodef{LIF}[LIF]{Leaky Integrate-and-Fire}
\acrodef{SNN}[SNN]{spiking neural network}
\acrodef{DoA}[DoA]{degrees of actuation}

\maketitle

\begin{abstract}
Regression-based decoding of continuous movements is essential for \acp{HMI}, such as prosthetic control. This study explores a feature-based approach to encoding \ac{sEMG} signals, focusing on the role of variability in neural-inspired population encoding. By employing heterogeneous populations of \ac{LIF} neurons with varying sizes and diverse parameter distributions, we investigate how population size and variability in encoding parameters, such as membrane time constants and thresholds, influence decoding performance. Using a simple linear readout, we demonstrate that variability improves robustness and generalizability compared to single-neuron encoders. These findings emphasize the importance of optimizing variability and population size for efficient and scalable regression tasks in \acp{SNN}, paving the way for robust, low-power \ac{HMI} implementations.
\end{abstract}


\section{Introduction}
\label{sec:intro}
Tracking finger movements accurately and efficiently is essential for numerous applications, including \acfp{HMI} such as prosthetic hands, smart home systems, and virtual reality environments~\cite{Wang2024} where decoding finger movements in real-time is essential for seamless interaction. A common approach involves using \acf{sEMG} signals, which non-invasively capture the electrical activity of muscles from the skin, and mapping them to finger movements. This task often requires decoding continuous, time-varying patterns, making it more complex than discrete gesture recognition~\cite{Sun_etal23}.

Regression-based approaches inherently demand precise modeling of temporal dynamics. Deep learning methods, such as recurrent or temporal convolutional networks, have been employed to address these challenges and achieve high accuracy~\cite{zanghieri_regression, burrello2022bioformers, leroux2023online}. However, their significant computational limits their applicability for lightweight, embedded systems critical for edge-computing applications like wearable prosthetics and low-power virtual reality devices~\cite{Zhai_etal17}.

Feature-based methods offer a promising alternative, focusing on extracting meaningful information from \ac{sEMG} while maintaining generalizability. Unlike handcrafted features, which are application-specific, automated feature extraction methods provide an adaptable solution. Furthermore, integrating these approaches with event-based processing, particularly neuromorphic or neuromorphic-compatible designs, can lead to significantly lower power consumption, making them ideal for embedded and energy-constrained environments~\cite{Covi_etal21, Donati_Indiveri2023, Indiveri_Liu2015}.

One feature-based approach that has shown promise involves encoding the power of \ac{sEMG} signals within specific frequency bands~\cite{zanghieri2023event, Baracat_etal24}. This method leverages spiking-based feature extraction, which is particularly suited for integration with \acp{SNN}. By converting \ac{sEMG} signals into spike-based representations, it aligns seamlessly with the event-driven nature of \acfp{SNN}.
By focusing on extracting relevant features from the signal, this technique simplifies the decoding process by linearizing the relationship between \ac{sEMG} signals and finger movements~\cite{Baracat_etal24}. This simplification enables the use of a straightforward linear regression model, avoiding the need for computationally intensive algorithms. However, earlier implementations of this feature-based method have exploited ideal neuron models, relying on single neurons with fixed parameters that are carefully optimized for specific signals~\cite{zanghieri2023event, zanghieri2024event}. While this approach is effective, its deterministic nature does not fully exploit the computational advantages that variability can offer, particularly in enhancing robustness and generalizability across varying signal conditions.

In neuroscience, variability is a critical feature rather than a limitation, enhancing the robustness and adaptability of neural computations~\cite{Marder_etal06}. Neural circuits exhibit significant variability in properties such as firing thresholds, membrane time constants, and synaptic weights, which allows populations of neurons to encode information more effectively and adapt to diverse conditions~\cite{Faisal_etal08}. This variability improves generalization, enhances resilience to noise, and enables neural circuits to flexibly respond to changing environments. 
\begin{figure*}[ht!]
    \centering 
    \includegraphics[width=.9\textwidth]{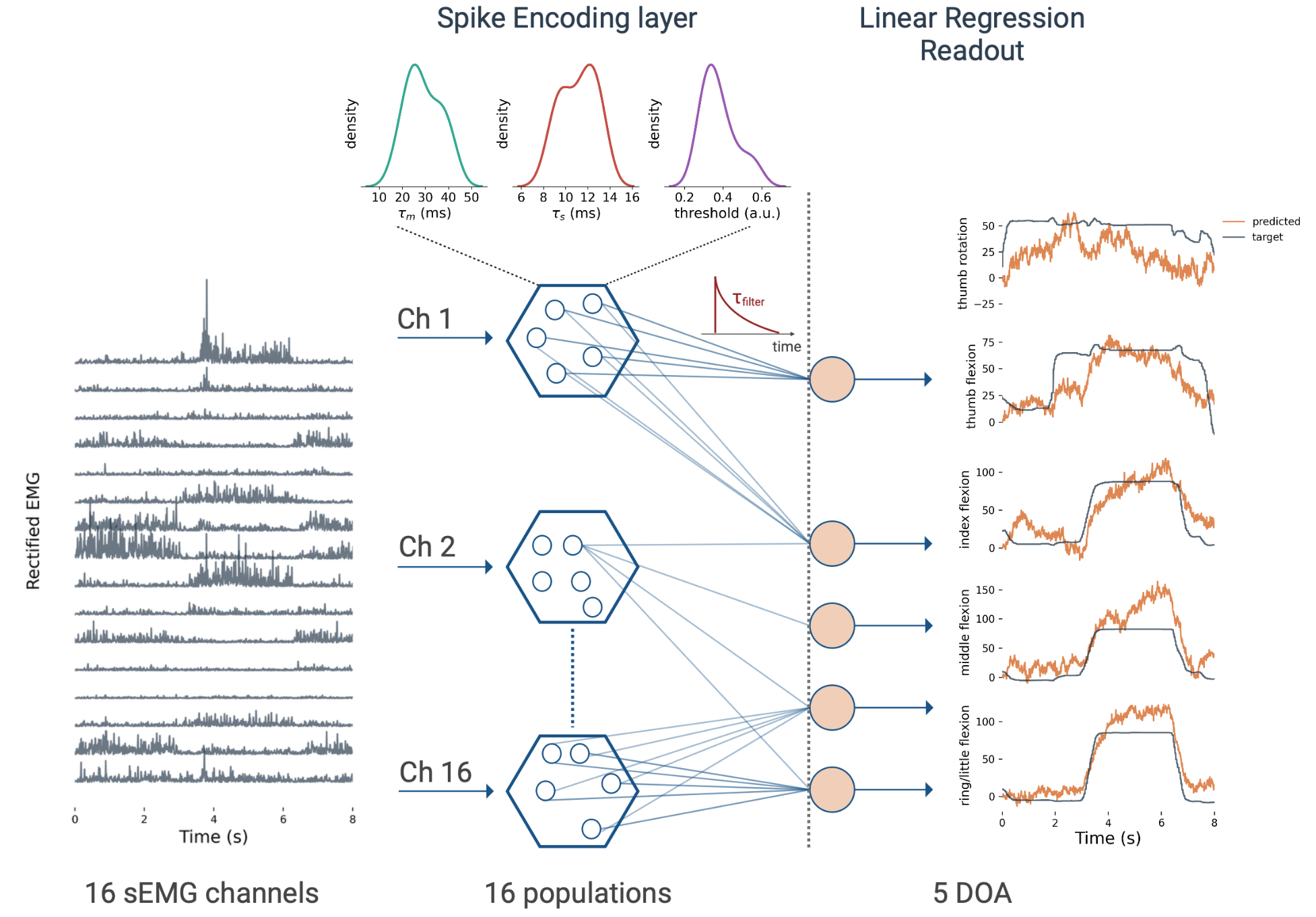}
    \caption{Network architecture. Each \ac{sEMG} channel is full-waved rectified and amplified by a fixed synaptic weight with a time constant ($\tau_{syn}$) before being injected into the corresponding population of heterogeneous \ac{LIF} neurons. Neurons are assigned membrane time constants ($\tau_{m}$) and thresholds sampled from normal distributions with 20\% variability. The encoded spiking activity is smoothened with an exponential kernel and decoded using a linear regression to predict the position of the 5 \acp{DoA}. An example input \ac{sEMG} channels and output predictions are shown for the cylindrical grip gesture (repetition 2 of subject 4) using an encoding population with 16 neurons per population,  $\tau_{m} = 30 \;ms$, $\tau_{syn} = 10 \;ms$ , and a threshold of 0.4 (a.u.).}
    \label{fig:net_architecture}
\end{figure*}

Building on this idea, we propose a novel approach that incorporates variability into the feature extraction process for decoding finger movements. Instead of relying on a single neuron with fixed parameters, we use a population of \ac{LIF} neurons, with properties drawn from Gaussian distributions. By introducing variability in parameters such as firing thresholds and membrane time constants, this method mimics the diversity observed in biological neural circuits. By systematically varying the mean of these distributions and the population size, we investigate the effect of variability on decoding performance.

This work demonstrates that variability, when appropriately implemented, enhances the computational capabilities of feature-based approaches. By leveraging diversity in neuron parameters, the proposed method achieves robust and efficient feature extraction, enabling lightweight, low-power solutions ideal for real-time, embedded applications. Furthermore, the method aligns with the principles of neuromorphic hardware, offering a scalable and biologically-inspired framework for decoding motor signals in human-machine interfaces.

\section{Methods}
\label{sec:methods}
To ensure robust evaluation of finger kinematic decoding from \ac{sEMG} signals, we systematically explored encoding parameters on a validated dataset, the NinaPro DB8 dataset~\cite{Krasoulis_etal19}, and metrics. This study focuses on systematically assessing the role of encoding population parameters and their ability to generalize across subjects while leveraging advanced \ac{SNN} simulations. The overall system is depicted in Figure~\ref{fig:net_architecture}.

\subsection{Dataset: NinaPro Database 8}
\label{ssec:dataset}
In our experiments, we used the publicly-available NinaPro DB8 dataset~\cite{Krasoulis_etal19}, a widely recognized benchmark for decoding finger positions from \ac{sEMG}.
The dataset consists of recordings from 12 participants (10 able-bodied individuals and 2 amputees) performing nine distinct gestures spanning individual finger movements and combinations of finger movements. Forearm muscle activity was recorded using an armband with 16 \ac{sEMG} electrodes, while precise finger movements were simultaneously captured using a CyberGlove 2 equipped with 18 sensors. Both \ac{sEMG} and glove data were upsampled to 2 kHz and post-synchronized. The 18 sensors of the CyberGlove 2 were subsequently linearly mapped to 5 \acp{DoA} of a prosthetic hand (see~\cite{Krasoulis_etal19} for more details). The final \acp{DoA} included the flexion/extension of the thumb, index, middle, and combined ring/little fingers, as well as thumb opposition.

\subsection{Network Architecture}
\label{ssec:network}
Figure~\ref{fig:net_architecture} illustrates the proposed network architecture, which consists of two layers: a spike-encoding layer for extracting features from the full-wave rectified \ac{sEMG} signals and a rate-based readout layer for decoding finger positions. The rationale for employing a rate-based readout is twofold: first, to enable a direct comparison with existing rate-based approaches, isolating the effect of population encoding from the intricacies of learning in \acp{SNN}~\cite{zanghieri2023event, zanghieri2024event}; and second, to establish a baseline for future benchmarking against fully spiking implementations.

The spike-encoding layer converts \ac{sEMG} features into spike train patterns using a power-based method inspired by the cochlea’s signal processing pipeline~~\cite{liu_siliconcochlea_1, zanghieri2023event, zanghieri2024event}. Initially, signals are filtered in one or more frequency bands, followed by full-wave rectification to preserve the complete information within the signal.  The rectified signals are then transmitted to \ac{LIF} neurons through synapses with fixed weights, where they are integrated by the neurons’ membrane potential, encoding the information into spike patterns. In this work, we used a single frequency band between 5 Hz - 500 Hz and focused on examining the impact of introducing variability in synaptic parameters (i.e. synaptic time constant, $\tau_{syn}$) and neuronal parameters (i.e. membrane time constant, $\tau_m$ and neuron threshold) on \ac{sEMG} encoding.

The 5 \acp{DoA} are decoded from the extracted features in the rate-based readout layer. The spike trains from all encoding neurons are processed by an exponential decay kernel with a fixed time constant ($\tau_{filt}= 200 \;ms$ ). This kernel is implemented using leaky integrator neurons, each connected one-to-one with a neuron in the encoding layer. Finally, we fit a multi-output linear regression model to map the filtered spike trains into 5 finger positions. 

We used snnTorch~\cite{eshraghian_training_2023} to simulate proposed network, with a simulation timestep ($\Delta t$) set to 10 ms. This choice of $\Delta t$ was made to expedite the simulation time. Notably, we did not notice any significant differences with smaller $\Delta t$.

\subsection{Encoding population parameters}
\label{ssec:encoding}
To comprehensively assess the impact of variability, we systematically varied the mean values of neuronal parameters within the encoding population, including the membrane time constant, threshold, and synaptic time constants, alongside the population size. Given that different population sizes may require distinct parameter tuning, our goal was to identify parameter regimes where the encoded \ac{sEMG} features effectively support the decoding of finger positions. The parameter sweep included  $\tau_{syn}$ values of [2 ms, 6 ms, 8 ms, 10 ms],  $\tau_m$  values of [10 ms, 20 ms, 30 ms, 40 ms], neuron threshold values of [0.4, 0.5, 0.6] (a.u.), and population sizes of [1 ,8, 16, 32, 64] neurons per population. This sweep was conducted for 4 out of the 12 subjects in the dataset.

\begin{figure}[ht!]
   \includegraphics[width=1\linewidth]{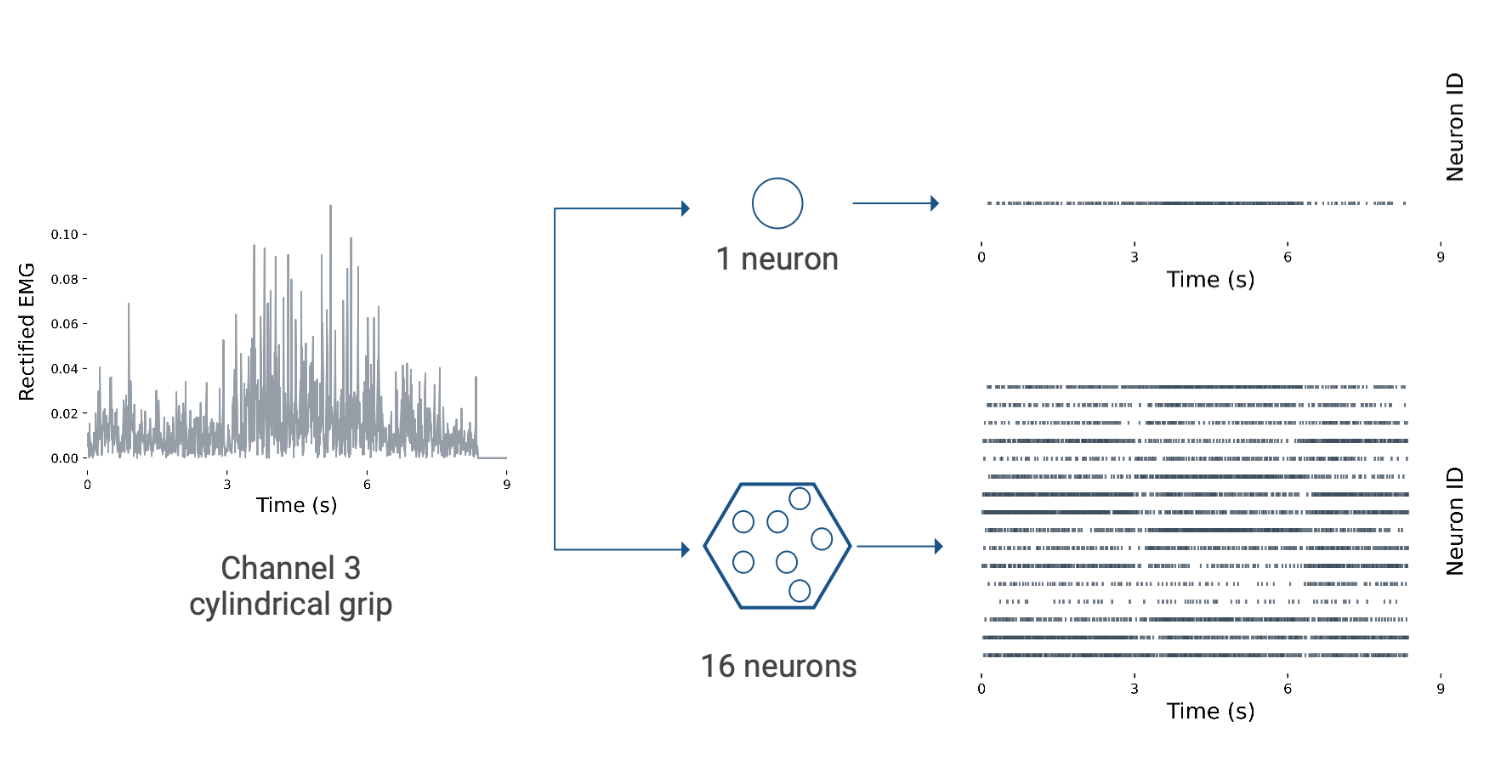}
\caption{A representative \ac{sEMG} signal from Channel 3 during a cylindrical grip task, encoded using two different approaches: (1) a single neuron encoding scheme (top right) and (2) a population encoding scheme with 16 neurons (bottom right). The single neuron approach produces a sparse spike train, while the population-based encoding generates richer, distributed spike activity across multiple neurons, highlighting the added representational capacity of a population for capturing signal dynamics.}
\label{fig:encoding_population_size}
\end{figure}

Figure~\ref{fig:encoding_population_size} illustrates two encoding strategies applied to a representative \ac{sEMG} signal from Channel 3 during a cylindrical grip task. In the single-neuron encoding scheme (top right), the output is a sparse spike train. In contrast, the 16-neuron population encoding scheme (bottom right) exemplifies population behavior by generating diverse spike trains, despite receiving the same input. Each neuron responds to slightly different features of the input signal. This demonstrates the increased representational power of the population approach, as the distributed activity across multiple neurons captures more nuanced aspects of the signal's temporal characteristics.
\vspace{-0.8cm}
\begin{figure*}[ht!]
    \begin{subfigure}[t]{0.5\textwidth}
   \includegraphics[width=1\linewidth]{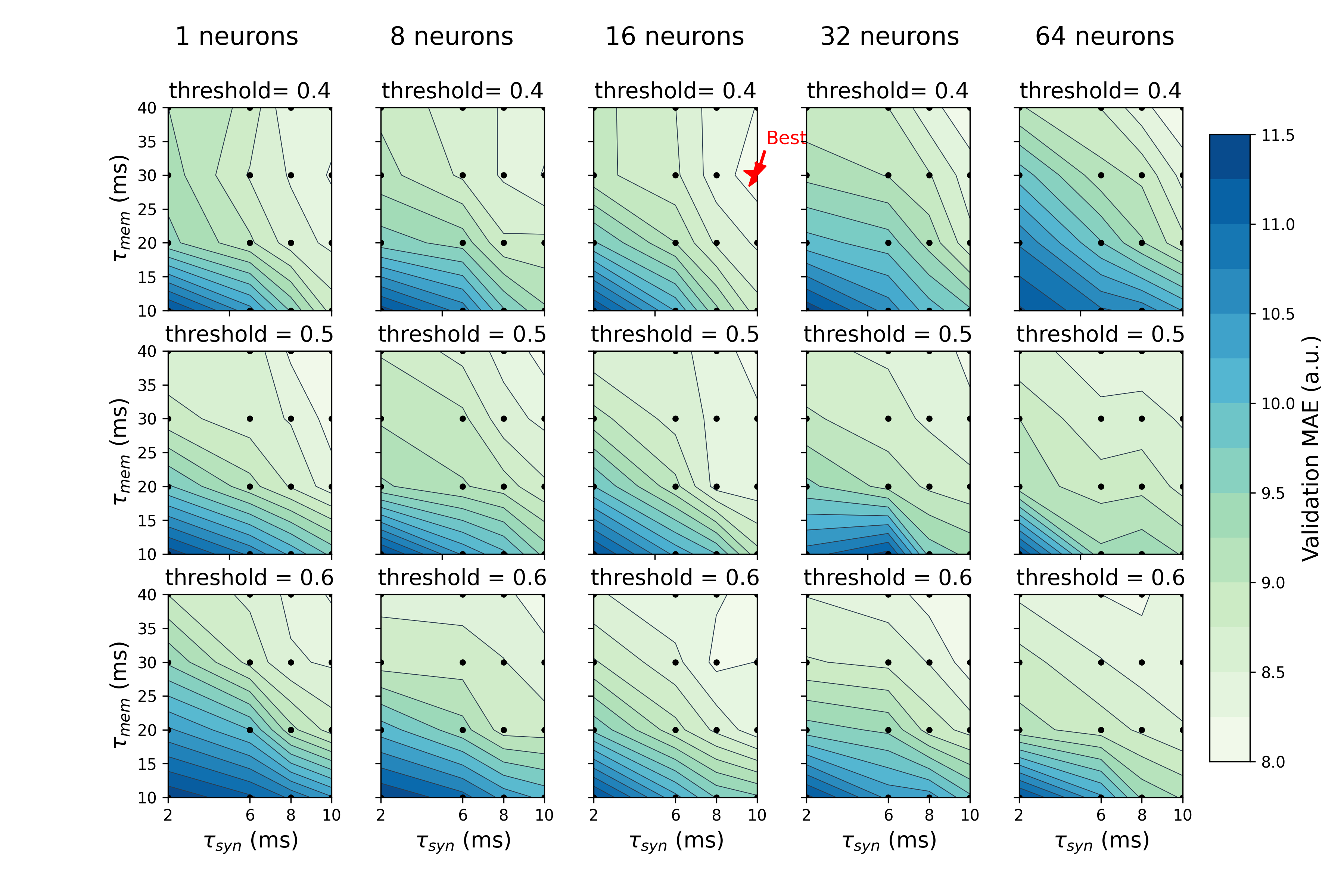}
   \caption{}
   \end{subfigure}
   \hfill
    \begin{subfigure}[t]{0.5\textwidth}
   \includegraphics[width=1\linewidth]{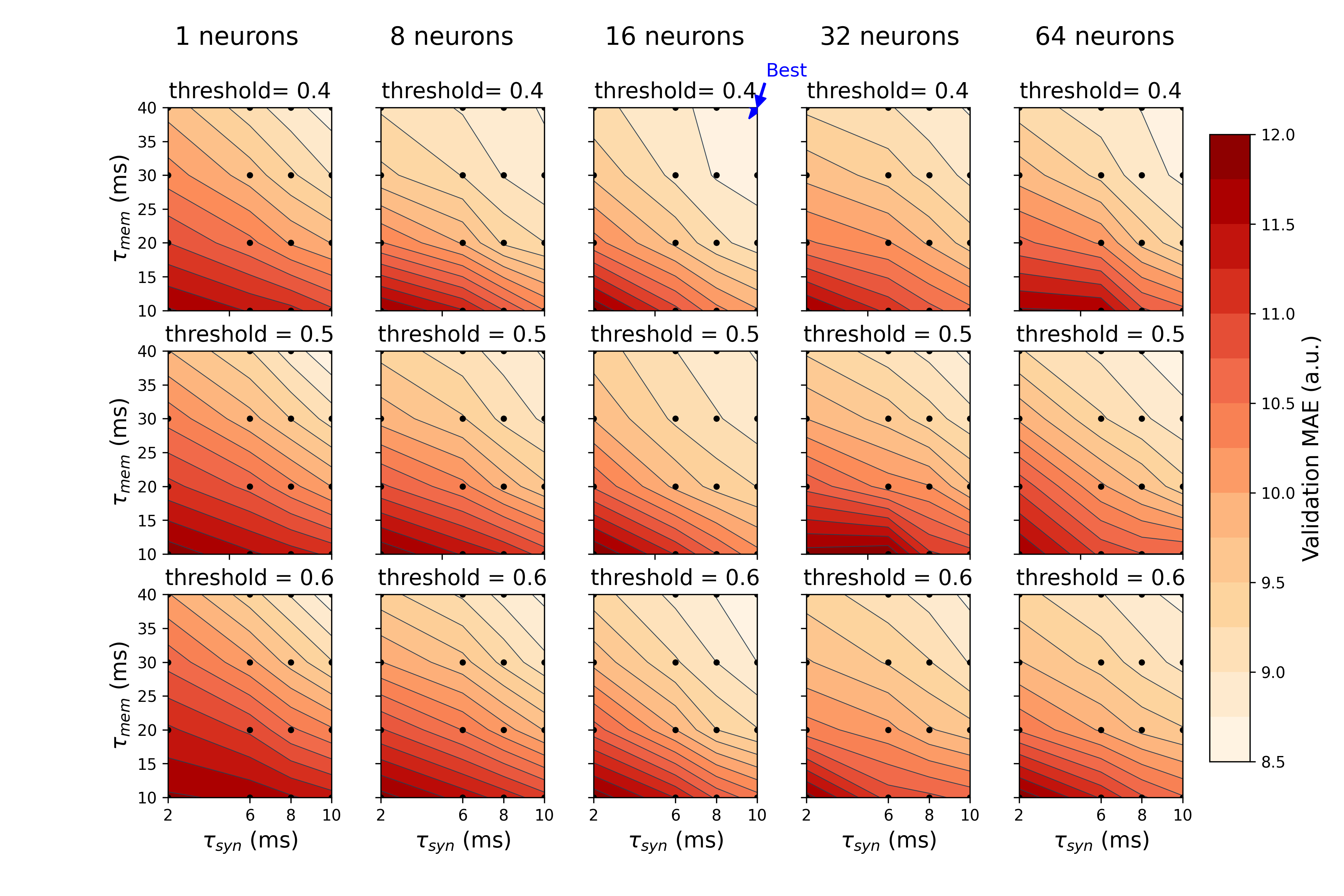}
   \caption{}
   \end{subfigure}
\caption{Effect of neuron parameters and synaptic time constants of the encoding populations on the validation dataset \ac{MAE} scores. (a) Results for a single representative subject (Subject 4). (b) Averaged \ac{MAE} scores across four subjects for the same parameters. In all cases, the exponential filter time constant, $\tau_{filt}$ is fixed to 200 ms. The best results are highlighted in both plots.}
\label{fig:encoding_parameters}
\end{figure*}
\begin{figure}[ht!]
   \includegraphics[width=\linewidth]{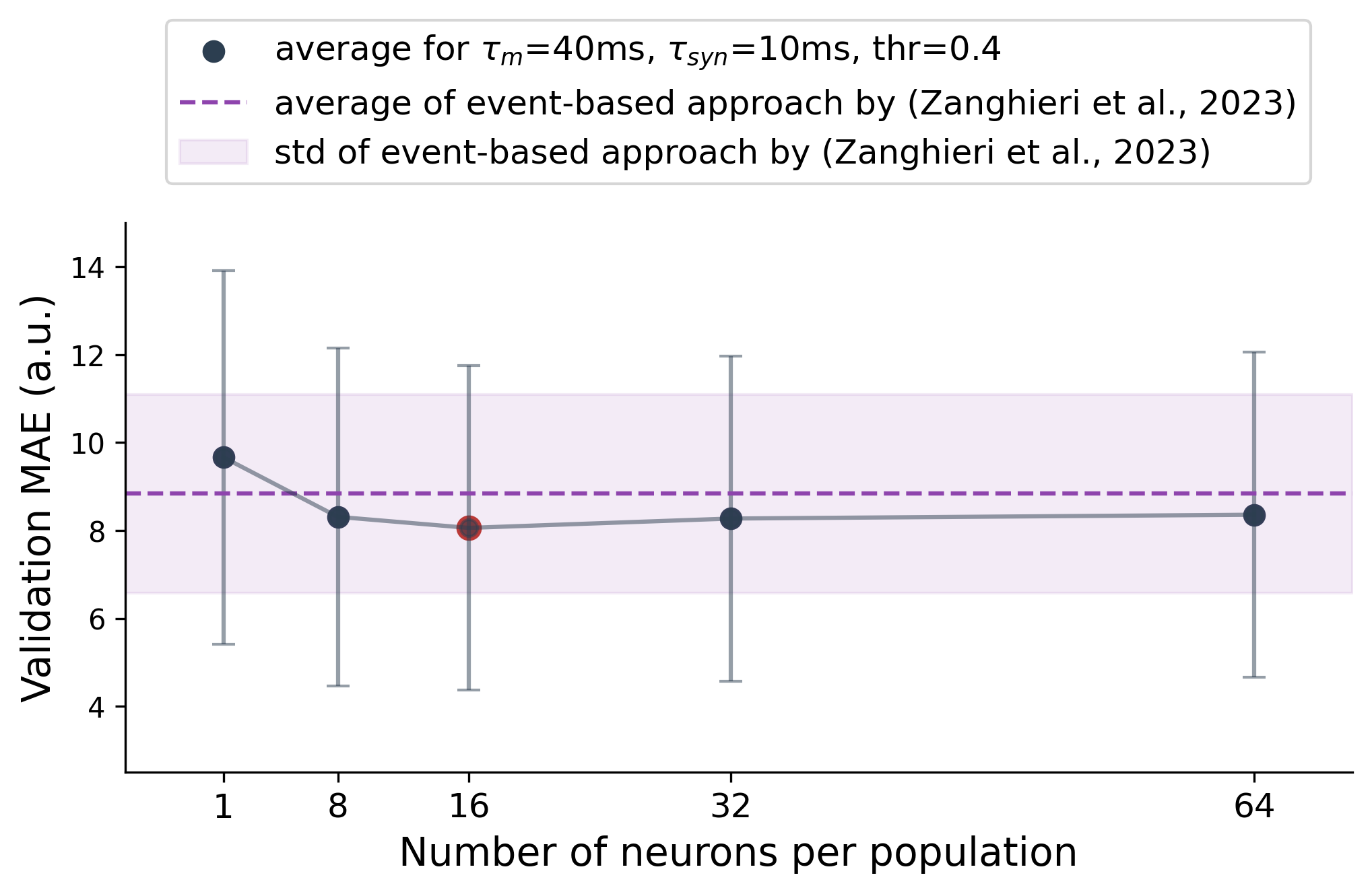}
\caption{Effect of encoding population size on decoding performance. Results are averaged across 8 subjects, with all encoding populations using the same parameters. The dashed line and shaded region represent the average and standard deviation, respectively, of the event-based approach reported in~\cite{zanghieri2023event} with an average \ac{MAE} $8.84 \pm 2.26$.}
\label{fig:avarage_mean}
\end{figure}

\subsection{Evaluation Metrics}
\label{ssec:metrics}
We evaluate regression performance using the \ac{MAE}, defined in degrees as:
\begin{equation}\label{eqn:mae}
\text{MAE} = 
\frac{1}{N_\text{infer} N_\text{DoA}} \sum_{i = 1}^{N_\text{infer}}
{\lVert \hat{\mathbf{y}}_i - \mathbf{y}_i \rVert}_1
\end{equation}
where $\mathbf{y}_i, \hat{\mathbf{y}}i \in \mathbb{R}^5$ represent the multivariate ground truth and estimated values, respectively, in degrees for the $i$-th inference, ${\lVert \cdot \rVert}1$ denotes the $L_1$-norm, $N\text{DoA} = 5$ is the number of  \acp{DoA}, and $N\text{infer}$ is the total number of inferences in Session 2 for each subject, with one inference occurring every 10 ms ($\Delta t$ = 10 ms).

The \ac{MAE} is a reliable and interpretable metric. It directly reflects the magnitude of errors in the same unit as the target joint angles, making it well-suited for assessing the accuracy of finger movement estimation or prosthetic control. Unlike second-order metrics such as \ac{RMSE} or $R^2$, which can disproportionately amplify the effect of large deviations, \ac{MAE} provides a balanced evaluation that is robust against transient decoding errors or noise commonly present in \ac{sEMG} signals. 

In this study, we calculate \ac{MAE} as described in Equation~\ref{eqn:mae}, averaging the error over time (across all movement types and repetitions), \acp{DoA}, and the 12 participants included in the analysis. This comprehensive averaging not only accounts for temporal variations in \ac{sEMG} signals but also captures inter-subject differences, providing a thorough evaluation of system performance. By enabling direct comparison with prior studies~\cite{zanghieri2024event}, this approach situates our findings within the broader context of \ac{sEMG}-based decoding research for \acp{HMI}.

\section{Experimental Results}
\subsection{Effect of Heterogeneous Encoding Neurons}
\label{ssec:heterogeneous}
Figure~\ref{fig:encoding_parameters}(a) presents the effect of neuron parameters and synaptic time constants on the \ac{MAE} scores for the validation dataset, using Subject 4 as a representative example. The rows indicate different neuron thresholds (0.4, 0.5, and 0.6 a.u.), while the columns represent varying sizes of the encoding population (1, 8, 16, 32, and 64 neurons). Each subplot explores the interaction between the synaptic time constant $\tau_{syn}$ on the x-axis and the membrane time constant $\tau_{m}$ on the y-axis, with the exponential filter time constant $\tau_{filt}$ fixed to 200 ms.

The contour lines and color gradient depict the \ac{MAE} scores, where lighter shades of blue correspond to lower \ac{MAE} values, indicating better performance. The results show that a population size of 16 neurons generally improves performance. Additionally, specific combinations of $\tau_{m}$, $\tau_{syn}$ and neuron threshold yield optimal performance ($\tau_{m}=30 ms$ and $\tau_{syn}=10ms$, threshold=0.4 a.u. for Subject 4), suggesting the importance of fine-tuning these parameters to capture a good representation of the input signal.

Notably, higher neuron thresholds tend to increase \ac{MAE}, as they produce sparser representations of the input signal. Furthermore, an evident interplay between neuron threshold and population size is observed. For instance, in a population of 64 neurons, a lower threshold restricts the network’s operating regime to a narrower parameter range, allowing fewer combinations of synaptic and membrane time constants to achieve good performance. 

\subsection{Generalization across Subjects}
\label{ssec:across}
Figure~\ref{fig:encoding_parameters}(b) figure presents the averaged validation \ac{MAE} scores across four subjects (Subjects 1-4) for the same set of neuron parameters and synaptic time constants shown in Figure~\ref{fig:encoding_parameters}(a). 
Compared to the first figure, which depicted results for a single subject, this figure provides a broader view of the parameter space by averaging across multiple subjects, reducing the impact of intrasubject variability. The first figure highlights this variability, showing subject-specific differences in optimal parameter combinations. By averaging the results, this plot captures general trends and helps identify robust parameter settings that perform well across individuals.

The selected parameter set, highlighted in this figure, represents the optimal combination of  $\tau_{m}$,  $\tau_{syn}$, threshold, and neuron population size based on this averaged data (using subset of Subjects). This chosen configuration is then used consistently across all 12 subjects in subsequent analyses, without any further re-tuning, ensuring a standardized approach while maintaining strong decoding performance. 

Figure~\ref{fig:avarage_mean} illustrates the validation MAE scores averaged across remaining unseen subjects (Subjects 5-12), including 2 amputees, using the optimal parameters identified from Figure 2(b) ($\tau_{m}$ = 40ms and $\tau_{syn}$ = 10ms,threshold=0.4). The Figure shows the \ac{MAE} over the population size, with error-bars reporting $\pm  \sigma\big(\rm{MAE}\big)$ estimated across 8 subjects. The results demonstrate that increasing the population size from 1 to 16 neurons improves decoding performance from a mean \ac{MAE} $9.67 \pm	4.25$ to $8.06 \pm 3.69$. This improvement highlights the benefit of enhanced neuronal diversity in the encoding layer, which allows for more robust feature extraction. However, beyond 16 neurons, the performance begins to plateau and slightly degrade as the population size increases further to 32 and 64 neurons. This result, together with results in Figure~\ref{fig:encoding_parameters}, show that while increasing the population size initially improves decoding performance by enhancing neuronal diversity, excessively large populations can degrade accuracy. This may occur because, as the number of neurons increases but the variance remains fixed, the system might suffer from an over-representation of features. Alternatively, the regression layer might over-fit due to the higher dimensionality introduced by the larger feature set (i.e encoding neurons).

Furthermore, the proposed approach is compared with previous results~\cite{zanghieri2023event}, represented by the dashed line and shaded area in the figure, which indicate the mean and standard deviation estimated across different subjects' performance, respectively. The findings reveal that the proposed method achieves comparable or superior performance, particularly at the optimal population size of 16 neurons.

These results underscore the importance of identifying an optimal population size that balances encoding diversity, neuronal variability, and computational efficiency while minimizing redundancy and over-fitting, highlighting the need for careful parameter tuning to achieve robust and efficient decoding in applications involving both able-bodied individuals and amputees.

\section{Conclusion}
\label{sec:conclusion}
This study highlights the importance of leveraging time-encoded features and controlled variability for accurate regression in \acp{SNN}. While variability improves performance compared to single-neuron populations, overly large populations reduce accuracy due to redundancy and noise. These findings underscore the need to optimize population size and variability, as well as to further develop time-encoded feature extraction methods, to fully exploit the capabilities of \acp{SNN} and pave the way for robust and efficient implementations in \ac{HMI} applications.

\section{Acknowledgment}
We gratefully acknowledge the support of the Royal Society through the International Exchanges Award IEC\textbackslash NSFC\textbackslash 223433, which facilitated collaboration and contributed significantly to the advancement of this work.

\printbibliography

@article{Wang2024,
  title={Application of Artificial Intelligence in Hand Gesture Recognition with Virtual Reality: Survey and Analysis of Hand Gesture Hardware Selection},
  author={Wang, Jindi},
  journal={arXiv preprint arXiv:2405.16264},
  year={2024}
}

@article{Sun_etal23,
  title={Continuous Gesture Recognition and Force Estimation using sEMG signal},
  author={Sun, Xuhui and Liu, Yinhua and Niu, Hequn},
  journal={IEEE Access},
  year={2023},
  publisher={IEEE}
}

@inproceedings{burrello2022bioformers,
  title={Bioformers: Embedding transformers for ultra-low power semg-based gesture recognition},
  author={Burrello, Alessio and Morghet, Francesco Bianco and Scherer, Moritz and Benatti, Simone and Benini, Luca and Macii, Enrico and Poncino, Massimo and Pagliari, Daniele Jahier},
  booktitle={2022 Design, Automation \& Test in Europe Conference \& Exhibition (DATE)},
  pages={1443--1448},
  year={2022},
  organization={IEEE}
}

@inproceedings{leroux2023online,
  title={Online transformers with spiking neurons for fast prosthetic hand control},
  author={Leroux, Nathan and Finkbeiner, Jan and Neftci, Emre},
  booktitle={2023 IEEE Biomedical Circuits and Systems Conference (BioCAS)},
  pages={1--6},
  year={2023},
  organization={IEEE}
}

@INPROCEEDINGS{zanghieri_regression,
  author={Zanghieri, M. and others},
  booktitle={IEEE COINS 2021},
  title={sEMG-based Regression of Hand Kinematics with Temporal Convolutional Networks on a Low-Power Edge Microcontroller}, 
  year={2021},
  doi={10.1109/COINS51742.2021.9524188}}

@article{Zhai_etal17,
  title={Self-recalibrating surface EMG pattern recognition for neuroprosthesis control based on convolutional neural network},
  author={Zhai, Xiaolong and Jelfs, Beth and Chan, Rosa HM and Tin, Chung},
  journal={Frontiers in neuroscience},
  volume={11},
  pages={379},
  year={2017},
  publisher={Frontiers Media SA}
}

@article{zanghieri2024event,
  title={Event-based Estimation of Hand Forces from High-Density Surface EMG on a Parallel Ultra-Low-Power Microcontroller},
  author={Zanghieri, Marcello and Rapa, Pierangelo Maria and Orlandi, Mattia and Donati, Elisa and Benini, Luca and Benatti, Simone},
  journal={IEEE Sensors Journal},
  year={2024},
  publisher={IEEE}
}

@inproceedings{zanghieri2023event,
  title={Event-based low-power and low-latency regression method for hand kinematics from surface EMG},
  author={Zanghieri, Marcello and Benatti, Simone and Benini, Luca and Donati, Elisa},
  booktitle={2023 9th International Workshop on Advances in Sensors and Interfaces (IWASI)},
  pages={293--298},
  year={2023},
  organization={IEEE}
}

@article{Donati_Indiveri2023,
  title={Neuromorphic bioelectronic medicine for nervous system interfaces: from neural computational primitives to medical applications},
  author={Donati, Elisa and Indiveri, Giacomo},
  journal={Progress in Biomedical Engineering},
  volume={5},
  number={1},
  pages={013002},
  year={2023},
  publisher={IOP Publishing}
}

@article{Indiveri_Liu2015,
  title={Memory and information processing in neuromorphic systems},
  author={Indiveri, Giacomo and Liu, Shih-Chii},
  journal={Proceedings of the IEEE},
  volume={103},
  number={8},
  pages={1379--1397},
  year={2015},
  publisher={IEEE}
}

@article{Covi_etal21,
  title={Adaptive extreme edge computing for wearable devices},
  author={Covi, Erika and Donati, Elisa and Liang, Xiangpeng and Kappel, David and Heidari, Hadi and Payvand, Melika and Wang, Wei},
  journal={Frontiers in Neuroscience},
  volume={15},
  pages={611300},
  year={2021},
  publisher={Frontiers Media SA}
}

@article{Marder_etal06,
  title={Variability, compensation and homeostasis in neuron and network function},
  author={Marder, Eve and Goaillard, Jean-Marc},
  journal={Nature Reviews Neuroscience},
  volume={7},
  number={7},
  pages={563--574},
  year={2006},
  publisher={Nature Publishing Group UK London}
}

@article{Faisal_etal08,
  title={Noise in the nervous system},
  author={Faisal, A Aldo and Selen, Luc PJ and Wolpert, Daniel M},
  journal={Nature reviews neuroscience},
  volume={9},
  number={4},
  pages={292--303},
  year={2008},
  publisher={Nature Publishing Group}
}

@article{Krasoulis_etal19,
  title={Effect of user practice on prosthetic finger control with an intuitive myoelectric decoder},
  author={Krasoulis, Agamemnon and Vijayakumar, Sethu and Nazarpour, Kianoush},
  journal={Frontiers in neuroscience},
  volume={13},
  pages={891},
  year={2019},
  publisher={Frontiers Media SA}
}

@article{Baracat_etal24,
  title={Decoding gestures from intraneural recordings of a transradial amputee using event-based processing},
  author={Baracat, Farah and Mazzoni, Alberto and Micera, Silvestro and Indiveri, Giacomo and Donati, Elisa},
  year={2024}
}

@INPROCEEDINGS{liu_siliconcochlea_1,
  author={Liu, S.-C. and others},
  booktitle={IEEE ISCAS 2010}, 
  title={Event-based 64-channel Binaural Silicon Cochlea with Q Enhancement Mechanisms}, 
  year={2010},
  volume={},
  number={},
  doi={10.1109/ISCAS.2010.5537164}}

@article{eshraghian_training_2023,
	title = {Training {Spiking} {Neural} {Networks} {Using} {Lessons} {From} {Deep} {Learning}},
	volume = {111},
	issn = {1558-2256},
	url = {https://ieeexplore.ieee.org/abstract/document/10242251},
	doi = {10.1109/JPROC.2023.3308088},
	number = {9},
	journal = {Proceedings of the IEEE},
	author = {Eshraghian, Jason K. and Ward, Max and Neftci, Emre O. and Wang, Xinxin and Lenz, Gregor and Dwivedi, Girish and Bennamoun, Mohammed and Jeong, Doo Seok and Lu, Wei D.},
	year = {2023},
	pages = {1016--1054},
}

\end{document}